# Innovation Systems as Patent Networks:

# The Netherlands, India and Nanotech

Wilfred Dolfsma[1] & Loet Leydesdorff[2]

**Abstract.** Research in the domain of 'Innovation Studies' has been claimed to allow for the study of how technology will develop in the future. Some suggest that the National and Sectoral Innovation Systems literature has become bogged down, however, into case studies of how specific institutions affect innovation in a specific country. A useful notion for policy makers in particular, Balzat & Hanusch (2004) argued that there is a need for NIS studies to develop complementary and also quantitative methods in order to generate new insights that are comparable across national borders. We use data for patents granted by the World Intellectual Property Organization (WIPO) to map innovation systems. Groupings of patents into primary and secondary classes (co-classification) can be used as relational indicators. Knowledge from one class may be more easily used in another class when a co-classification relation exists. Using social network analysis, we map the co-classification of patents among classes and thus indicate what characterizes an innovation system.

A main contribution of this paper is methodological, adding to the repertoire of methods NIS studies use and using information from patents in a different way. Policy makers may also find benefits in the social network analysis of the complete set of patents granted by the WIPO to firms and individuals in a country. Social network analysis indicates what innovation activity occurs in a countries and which fields of technology are likely to give rise to innovative products in the near future. We offer such analysis for the Dutch and Indian Innovation Systems. This social network analysis could also be done for a Sector Innovation System, and we do so for

[1] Corresponding author: Universiteit van Groningen, School of Economics and Business, PO Box 800, 9700 AV Groningen, the Netherlands, ph. +31-50-363 2789, fax +31-50-7110, w.a.dolfsma@rug.nl

[2] Amsterdam School of Communications Research (ASCoR), Universiteit van Amsterdam, Kloveniersburgwal 48, 1012 CX Amsterdam, The Netherlands; loet@leydesdorff.net, http://www.leydesdorff.net

Nanotech to determine empirically the knowledge field relevant for this emerging scientific domain.

**Acknowledgements.** This paper was presented at a number of seminars and conferences. We would like to thank all participants at EAEPE conference (Rome), Innovatieplatform (The Hague), ISID (New Delhi) and UNAM (Mexico City). The usual disclaimer applies.

# Innovation Systems as Patent Networks

# The Netherlands, India and Nanotech

> "... there is much more information derivable from the patent documents than just simply their aggregated numbers in a particular year or for a particular firm." (Griliches 1990, p.1664)

Both academics and policy makers have found the National Innovation Systems literature to contribute useful insights. Some, however, believe that the National Innovation Systems – NIS – literature (Edquist 2004; Lundvall 1992; Nelson 1993) has become bogged down into case studies of how specific institutions affect innovation in a specific country. As Balzat & Hanusch (2004) argue: there is a need for NIS studies to develop complementary and also quantitative methods in order to generate new insights that are comparable across national borders. In this paper we use data on patents granted by the World Intellectual Property Organization (WIPO), a UN organization, to map innovation systems. Applying for a patent at the WIPO is relatively easy and cheap, and allows the applicant to both apply in relevant markets afterwards, and establish their position vis-à-vis competitors. Rather than taking counts of the number of patents granted, by field, company, region or country, which would meet with all the drawbacks that patents have as an indicator for innovation (Kleinknecht *et al*. 2002), we use different information that can be drawn from patents.

Patents are grouped into a primary class and secondary classes by patent examiners. Co-classification of a patent in two classes signifies a relation between these classes that is significant from the point of view of knowledge development and thus for studying a knowledge-based innovation system. Using social network analysis, one can map these co-classifications among patent classes and thus characterize a national innovation system. Such an analysis of a national innovation system focuses on what nation specific components and relationships in a system, each with their characteristics and attributes (Carlsson *et al*. 2002), actually produce. It does so in a manner that indicates the relations between knowledge fields as well as, to some degree, the nature of such relations as part of the larger (socio-cognitive) network configuration. In doing so, the results of an analysis of (national) innovation systems becomes amenable for comparison (between nations) (Liu & White 2001).

One may argue that this approach ignores the idiosyncracies of national systems, but one may as well argue that such an analysis may enable us to understand these idiosyncracies. In addition, the analysis can focus on regions as well as specific technological fields, which may allow for theoretical integration between NIS, Regional Innovation Systems (Cantner & Graf 2006) and Sector Innovation Systems (Dittrich & Duysters 2007; Storz 2008) where a similar network approach can be adopted. Analysis of industrial production in terms of input-output matrices adopts or could adopt similar methods (Lenocini and Montresor 2000; Lotti & Santarelli 2001; Verspagen 1997).

The empirical analysis of national innovation systems as patent networks may thus open possibilities for theoretical integration of NIS to adjoining fields of academic research. Yet, the main contribution of this paper may perhaps be methodological – as it adds to the repertoire of methods NIS studies use, but also as a different kind of information on patents is used. In addition, policy makers are interested in knowing how knowledge development in an innovation system is interrelated, and thus obtain an understanding of how production structures may evolve in the near future. It might also indicate which policy domains may emerge as important issues. For instance, using social network analysis of the complete set of 3,287 patents granted by the WIPO to Dutch firms and individuals in 2006, we find that biotech, pharmaceutical and chemical technology, with applications in food and medication may be overtaking the traditionally dominant position of electronics / computer technology. Given that these technological fields and their associated industries show high propensities to patent, the dependence of the Dutch NIS on patent law thus increases. We also perform such an analysis for emerging economy India, and separately for the technological domain of nanotechnology.

## 1. Innovation Systems Literature

In terms of both direction and success rate, innovation performance differs widely across firms and organizations more generally grouped by regions, sectors or specifically nations. At these aggregated levels, a system's approach has been popular since at least the early 1990s to understand emergent phenomena (Nelson 1993; Lundvall 1988). A National Innovation Systems approach assumes that differences among the innovation performance across countries are due to their specificities and idiosyncrasies that will not (simply) disappear due to market processes (Dolfsma *et al*. 2008). Players and other components of a system each have their characteristics, while relationships among them affect both how they will contribute to system outcomes as well as how each evolves over time (Carlsson *et al*. 2002).

An institutional perspective is often invoked (Liu & White 2001), claiming that both informal and formal institutions can be persistently different among countries in a way that affects innovation patterns and outcomes. Actors and networks are also referred to in this respect (Carlsson & Stankiewicz 1991), their workings perhaps best understood in terms of institutions and what they legitimately allow or prescribe (Bergek *et al*. 2008; Dolfsma & Verburg 2008]) In this respect, and contra Bergek *et al*. (2008), a system can be considered as more than an analytical abstract: one can formulate as an empirical question whether or not at the national level one can perceive coherence in the institutional structure with regard to innovation.[3] The outcomes of the largely unplanned workings of an innovation system may thus be difficult to predict precisely, but can be approached. The NIS approach is an attractive starting point because of its coherence and usefulness for policy.

Using the nation as units of analysis, authors often refer to its institutions that help create new knowledge, such as strong universities, an attractive climate for private research institutes, possibilities for migrant knowledge workers to enter a country and a patent system. Many such institutions also play a role in knowledge diffusion.[4] A well-functioning education system, people's attitude towards taking the risk of setting up a new firm, or a country's laws with regard to for instance bankruptcy are other institutions that play a role in knowledge diffusion.

While a useful approach in many ways, the promise of comparability across national systems has been largely unmet (Edquist 2004; Liu & White 2001). This may be due to the heterogeneous nature of the concept of NIS or its constituent parts (Bergek *et al*. 2008), but may as well be due to the empirical and case-study based approach taken as is evident in history-friendly analyses such as in Nelson (1993). Quite a few studies inspired by a national innovation system idea have focused on a limited number of institutions, or even a single one, to study their effect on innovation direction and performance. The advantage of the approach – an awareness of idiosyncrasies – may then become a drawback since from up close the differences between systems stand out more than the similarities. It can thus be challenging to point to causes for the differences or the similarities when comparing between systems. The choice of institutions and the choice regarding the aggregation level for analysis can differ substantially between studies, resulting in a situation where some believe that the approach has come to be stranded due to the case study approach adopted.

---

[3] Using a different indicator, Leydesdorff & Fritsch (2006) found that Germany cannot be considered integrated nationally as an innovation system, while the Netherlands can (Leydesdorff *et al.* 2006).

[4] The NIS literature may be more focused on knowledge creation, while knowledge diffusion may be of greater importance to the knowledge economy (Leydesdorff *et al*. 2006).

We concur with Balzat & Hanusch (2004) that it is possible to salvage a NIS analysis by developing additional, complementary approaches to the study of national innovation patterns. These include, but may not be restricted to, quantitative methods. One may have to sacrifice to some extent the attractive feature of a rich – or, as anthropologist Geertz (1973) called it, "thick" – description as one focuses less on the workings of a system rather than the outcome, but one potentially gains as comparability and rigor is enhanced.

We propose to use patent data to study the structural characteristics for the outcomes of an Innovation System. As a result of the interplay among actors within an innovation system, behaviour which is determined by extant institutions, both the direction of technological development as well as the robustness of that pattern emerge (Carlsson *et al*. 2002; Rip & Kemp 1998). While analysis of a particular NIS or of a particular set of players within a NIS allows for detailed analysis of the dynamics of a NIS over time (Storz 2008), the approach we opt for also allows for comparison over time of the way in which elements and functions of a NIS (Bergek *et al*. 2008; Liu & White 2001) produce innovations. This holds, we would argue, for both an analysis of national innovation systems as well as for an analysis of a sector or technological innovation system (Storz 2008; Malerba & Orsenigo 1997).

While patent data offers a quantitative measure, they are often used in a rather unimaginative way. Patents granted are aggregated to the level of firms, regions, sectors or countries to determine the respective significance of innovative activity for an entity at such an aggregated level. While patent data is shaped by institutions, and reflects information about applications that is the result of institutional configurations, they are not defined by institutions a priori. Institutions themselves are embedded in knowledge infrastructures, providing the technological opportunities that have to be interfaced with market positions and expected demand that agents can act upon.

From this perspective, patent data offer a vastly more informative source of information, for instance about actual or potential knowledge flows in a system. We analyse patents granted as sediment of substantive-technical efforts by actors to develop new knowledge or find new (non-obvious) applications for existing knowledge. Classes of patent applications, and particularly co-classifications, thus may be taken as an important indicator of a mutual knowledge basis within the boundaries of a system (Breschi *et al*. 2003; Leydesdorff 2008). The network of co-classifications for patents, drawing on a unified and harmonized database, thus indicates the workings of an innovation system such as NIS and its relevant institutions. This approach to the studying of innovation systems would seem to be comprehensive in the sense of using a full set of data that gives some indication of the relevant output on an innovation system. We suggest that this approach is complementary to other methodological approaches developed to analyse innovation systems.

## 2. Data and Method

***Patents.*** Patents have been a widely used type of data for innovation studies, in part because of their availability. Patents granted in the United States, for many sectors the most important single market, are easily downloadable from the USPTO website. Such US data may not be relevant for the characterization of, for instance, a European country (Criscuolo 2006; Leydesdorff 2004). Patent data as a measure of innovativeness of a country, sector or firm has more generally come under increased discussion. Patents as an output measure of innovation is problematic – many of them do not have any commercial value for firms (Kleinknecht *et al*. 2002; Carlsson *et al*. 2002). As a result, the propensity to patent differs widely across industries (Arundel & Kabla 1998).

Of all patents granted in the US, 55-75 percent lapse and become a part of the public domain through failure to pay maintenance fees; if litigation against a patent's validity is a sign of commercial value of that patent, the fact that only 1.5% of patents are litigated and only 0.1 percent litigated to trial does not bode well (Lemley & Shapiro 2005; Dolfsma 2008). Many patents thus are applied for only strategic reasons (Dolfsma 2011; Granstrand 2000; Griliches 1990). Small firms are thus reluctant to patent in the technological or economic (market) 'neighborhood' of patents held by large companies for fear of expensive law suits (Lanjouw & Schankermann 2004;,Lerner 1995). In addition to this, the propensity to patent differs between product and process innovations, and by sector (Arundel & Kabla 1998). Fewer process innovations are patented, since secrecy as an alternative appropriability measure is more feasible. Propensity to patent also differs by sector. While this may somehow bias a picture of a national innovation system that is otherwise relatively unbiased because of the rather objective information used and the systematic way in which to collect that information. Patents are becoming increasingly important as a way to protect innovative knowledge, which means that the approach to innovation systems research that we suggest in this paper will over time face fewer objections.

Patent law tries to find a balance between the public interest of stimulating development of new and the wide diffusion of existing knowledge, on the one hand, and the private interest of return on investment, on the other. The institution of patent law, then, both seeks to facilitate two functions of an innovation system: knowledge development and diffusion [7]. Whether or not it is true that patent law stimulates knowledge development and diffusion, to the same degree, remains an empirical issue (Dolfsma 2008; Lerner 2009) but innovation is believed to be stimulated if innovators have a legal right to exclusively exploit the results of innovative efforts.

However, the balance is struck differently in different countries (OECD 1997), and in noticeably different ways. An important characteristic of the US patent law, for instance, which makes it unique, concerns who is deemed to have the right in an invention: the one who first files a patent application with the patent office or the one who is able to prove he was the first to invent. In the US the administratively less tractable first-to-invent may claim the rights. Parties who have been first to invent may decide to come forward with their invention only after another party is granted a patent by the patent office. This can give rise to legal conflicts that could prevent or limit the commercial use of the knowledge involved (cf Bittlingmayer 1988). A first to file system of administrating patent applications seems to be an example of public interests of providing clarity outweighing the private interests of the first inventor. Also, the  by providing a degree of certainty to the innovator applying for a patent as she can browse the patents applied for or granted to determine whether the knowledge embodied in a patent is already legally protected. In the US inventors who had not applied for a patent may even challenge a patent granted if they can convincingly show that they had been earlier to invent. On the other hand, in the US applicant only needs to publish the information contained in a patent after the patent is granted, while publication is required in Europe when a patent is applied for.

As a patent application can be rejected, the inventor thus runs the risk of diffusing her knowledge without receiving the legal right to exclusive commercial exploitation in return. This clause in European patent legislation favors the public interest more. This obviously opens possibilities for strategic behavior for parties (Granstrand 2000). However this may be, the important methodological point to make is that the level of analysis chosen for our analysis is that of the knowledge field (Carlsson et al. 2002). Specification of the boundaries of a knowledge field is left in the hands of patent applicants and patent officers.

**WIPO Patents.** The World Intellectual Property Organization (WIPO; www.wipo.org), an organization residing under the United Nations based in Geneva, offers the possibility to easily and cheaply apply for a patent. WIPO staff assists in drafting the patent application, which means that expensive additional technical and legal services that might be required for a European or US application need not be hired. The application for a patent submitted at WIPO can subsequently be submitted in other countries or jurisdictions as well within a specific time frame if commercially attractive. The legal systems of other countries recognize WIPO applications technically and legally. In addition, WIPO patents are part of the ‘prior art’ that patent officers need to consult in case they receive an application from a different party on a related technical invention. Such an application may then have to be rejected, or can relatively easily and cheaply be challenged in court by the patentee of a WIPO patent. In addition, due to the the low cost of applying for a patent at the WIPO as well, a larger number of small firms and applications from low-income countries apply for a patent at WIPO.

WIPO patent protection is thus an accessible means to obtain legal protection for an invention that may have industrial applications. Especially for parties that lack financial means, this makes applying for a WIPO patent attractive. Relatively small firms and parties from developing or emerging economies may find applying for a WIPO patent particularly attractive. Such parties may also have defensive motives to apply for a patent, preventing others to file for a patent in the same area. The tendency for smaller firms to shy away from R&D deployment in areas where larger firms already have a patent position for fear of being sued by these firms may thus diminish (Lanjouw & Schankermann 2004).

Litigation in patent law, specifically in the US, has grown increasingly rife, where especially large firms reserve substantial funds to legally defend their patent position even if technically their position might not seem particularly strong. Much patenting, again in the US in particular, is thus of a offensively strategic nature (Lemley and_Shapiro 2005). Needless to say, the number of patents applied for has increased substantially in recent years, at the USPTO as well as at WIPO where strategic patenting is less dominant. We may conclude that, as a source of information about technological development and innovation, WIPO data are more valuable.

**2006 WIPO patent data.** Patent databases are a rich source of information (Griliches 1990). One can do more than count the number of patents for each country or firm. Each patent is, for instance, given a classification indicating its technological field. Based on an understanding of the substantive technical knowledge in the patent, it is determined in what technological domain or paradigm (Dosi 1982) it belongs and on what previous knowledge it draws. In a substantial number of cases patents draw on knowledge previously developed in different paradigms. In addition, patent examiners provide co-classifications as well. The classification and co-classification indicates actual or potential knowledge transfer between different technological fields (Verspagen 2006). The 1994 OECD Manual (OECD 1994, p.52) mentions patent co-classifications as a potential indicator of linkages among technologies. Based on the patents granted to a particular entity, it can be established how the knowledge base of the entity can be characterized empirically (Breschi *et al*. 2003; Engelsman & Van Raan 1994).

Together with the European Patent Office (EPO), a.o.,[5] WIPO invests substantial resources in developing the International Patent Classification (IPC). Currently the eighth edition is in use, using a 12-digit coding system covering some 70,000 patent classes. Because of the standardized nature of the data presented in the WIPO database, patents registered there are a good source for data on innovation. Comparison across countries, or an analysis of a specific sector across country boundaries is also possible using this data. However, patent data from WIPO are not perfect as an indicator. As noted earlier, the propensity to patent, for example, is known to differ substantially across sectors. Also, on average only 35% of product and 25% of process innovations are patented (Arundel & Kabla 1998). The propensity to patent product innovation ranges between 8 and 80 percent. Other means may be deployed, and be deemed more important, to protect a firm's intellectual property. Secrecy is one of these (Levin *et al*. 1987).

[5] Inpadoc in Vienna, Austria.

Controlling for the number of citations to a patent, sometimes advocated as a way to control for the quality of patents in an analysis, may not present a better picture of the importance of or value for a patent as, at least for patents granted by the European Patent Office (EPO), there is only a very tenuous relation between patent value and number of citations (Gambardella *et al*. 2008).[6] This may be due to the fact that many citations in an application are included at the behest of the patent officer (Griliches 1990, p.1689). Nelson (2009) concurs, suggesting that licenses of a patent to other firms are a better indicator. Patents that are not licensed to other firms might, however, be important for the knowledge base of an innovation system nonetheless. Given these considerations we believe that the concerns raised about the use of patents in the context of analysis of innovation systems (Carlsson et al 2002, p.241) are properly addressed when co-classification information from patent data are used only and have decided to use such data in our analysis.

***Network analysis.*** From a network point of view, whenever co-classification between patent classes occurs, the classes thus connected may be said to be connected or tied together. The more such co-classifications occur, the stronger the tie between the classes. We use classifications and co-classifications for patents to analyze the innovation system as a network of related technological classes. For any of the 132 countries where individuals or organizations are located that had been granted one of the 138,751 patents by the WIPO in 2006 a network analysis of the

[6] Using US data, Hall *et al*. (2005) find that the number of citations to patents correlates with the value of the firm holding the patents measured in terms of Tobin's Q. Since the US patent office, the USPTO is known to be quite lenient when granting patents and given the size and importance of the US market for most firms, the signal of owning a patent granted by the USPTO might need to be complemented with additional information. They find self-citations, i.e. citations to other patents owned by the same firm, in particular to correlate to the market value of the firm. For the purposes of the analysis here, the focus is less on the specific firm that owns patents, nor their monetary value, but rather the focus is how knowledge develops and may subsequently be exchanged in an innovation system.

National Innovation Systems as a patent network can be prepared. We observe a spread in terms of the number of patents granted to parties in countries from 2 (Brunei Darussalam, Burundi, Cayman Islands, Côte D'Ivoire, Guatamala, San Marino, Seychelles, Tuvalu, Uzbekistan, Virgin Islands) to 48,190 (USA). If the number of patents granted to parties in a system is very low, a patent network may not present a useful view of an innovation system.

We analyze at the 4-digit level patent data. Circles or nodes in these figures are patent classes. Size of the nodes could be drawn such that they reflect the number of patents which have received such a (co-) classification. If we would have opted to do this, it would reflect a raw count of the number of patents in such a class. Given that our aim is to provide indication of the way in which technological domains or paradigms relate, we focus on the ties connecting the nodes. Lines between classes indicate co-classication, while thickness of lines indicates the number of co-classifications. The thicker the lines the more often co-classifications occur in the data. Thickness of the lines reflect something different from a raw patent count. While the number of patents in a class applied for by actors in a country reflects the research strengths of that country's innovation system in a way, we believe that the relations between classes provide insights that are complementary. Since clustering of research strengths can be easily traced and given that science increasingly develops where scientists from different disciplines interact and collaborate (Wuchty et al 2007), we believe that the insights provided by the networks of co-classified patents offer a fuller understanding. As these databases containing information of patents granted can be publically searched, this means more knowledge is actually or potentially exchanged between classes. WIPO recognizes 624 different major patent classes. The average number of classifications per patent is 2.4.

Social Network Analysis allows one to construct a visual presentation for the National Innovation System (De Nooy *et al*. 2005). To enhance readibility, weaker relations between classes can be excluded from a picture by imposing a threshold of a minimum number of co-classifications before a relation is included in a figure (either by using a *k*-core or the very much similar *m*-slice threshold level to trace clusters or cliques of more strongly interconnected sets of patent classes; cf. De Nooy *et al*. 2005).

## 3. Innovation Systems

To give indication of the usefulness of the approach suggested, in what follows, we compare two national innovation systems. For purposes of interpretation of the results we take the Netherlands as a developed country and India as an emerging country. The Netherlands is a highly developed economy that also turns out to be a mature innovation system. On the other hand, India is both an emerging economy as well as an emerging innovation system. One element which shows a relatively strong presence in India is Nanotechnology. The sector innovation system for nanotechnology, that we also present to indicate the potential of the kind of analysis we propose, shows a regime in development.

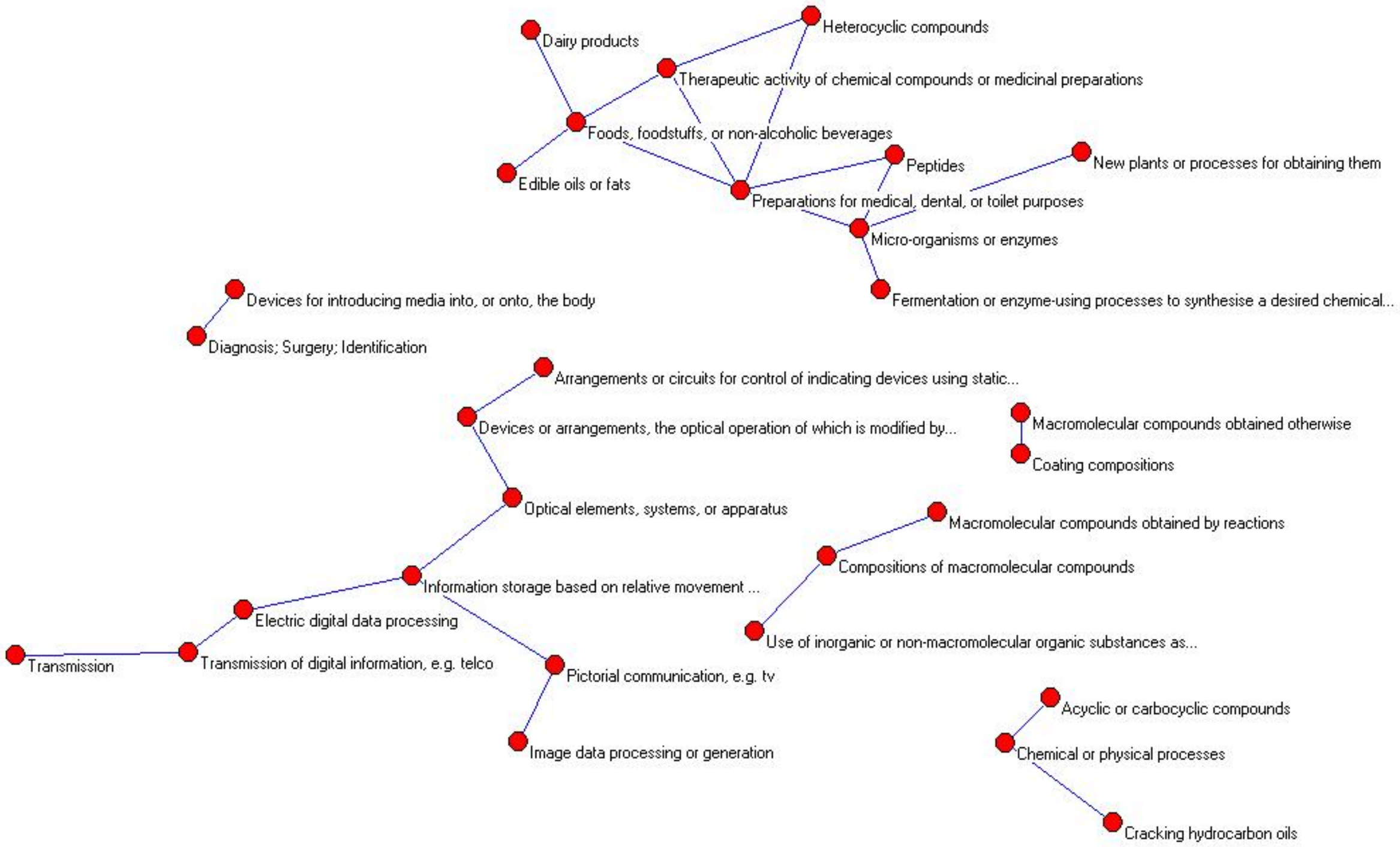

Figure 1: Patent classification categories and co-classification relations; core network for internationally registered Dutch patents (*k*-core = 10; 2006; *N* = 3,287). Source: WIPO.

**The Dutch Innovation System**. A factor analysis to find out if a cluster of patent classes among the 3,287 patents granted to parties in the Netherlands can be determined based on the extent to which they co-classify (available upon request from the authors) provides little clues as to which patent classes may be combined to form factors that help explain variance in the data. The factor analysis is explored as well at the 3-digit aggregation level. Only a small fraction of the variance (11.8%) is explained clustering 9 factors. This indicates that the Dutch innovation system is quite dispersed and broadly based, which may be a sign of its relative maturity.

Social Network Analysis offers as a highly attractive and informative possibility the option of visualization. Given the visualization of Figure 1, it is clear that the innovation complex around Eindhoven and the north of the Limburg province is strongly present. Some 50% of R&D formally spent by Dutch firms is spent in these two NUTS 3 regions (Leydesdorff *et al*. 2006). This is the region where electronics, computer and information processing technology, and optics cluster related to such firms as Philips, Océ, ASML, and supplying firms. A second large cluster, however, is that of chemical technology, biotech, and pharmaceutical technology, especially with applications in medication and functional foods. Even though firms in these sectors are more likely to apply for patents in case of an innovation than firms in different sectors, the cluster is larger and more closely knit than expected. This interrelated set of classes (chemical technology, biotech, and pharmaceutical technology, especially with applications in medication and functional foods) is not generally recognized as important in the Dutch innovation cluster, and its link to clusters that are recognized as key may go largely unnoticed (Ministry of Economic Affairs 2005; Innovatieplatform 2004, 2009). It is certainly a relatively younger cluster in terms of innovation focus, while the link to functional food indicates a relation to an traditional strength of the Dutch economy. A new high-tech stronghold seems to be developing.

In addition to these larger clusters of techno-economic activities, smaller clusters can be appreciated in the visualization. Chemical technology related to application in paints is visible in the network representation of the Dutch NIS despite the high (*k*-core) threshold applied. Multinational companies such as DSM and AkzoNobel are active here. So is oil refining, with companies such as Royal Dutch/Shell, despite the focus on process innovation and thus a lower propensity to patent if only for this reason in this mature industry. Given the recent change in strategic emphasis of the Dutch industrial behemoth Philips, the cluster indicating medical diagnostic equipment may change too, and possibly move in the direction of the electronics / computing cluster. Packaging, for example of food stuffs, is on the verge of being included. It is a sector that develops new products that year-upon-year are perceived by experts to be highly innovative and valuable. Consumers have been impressed more by the functional food mostly by dairy industrialists such as Campina and Friesland Foods (recently merged to form FrieslandCampina). Despite its small size and high population density, the Netherlands remains the second largest exporter of agricultural produce, to some extend undoubtedly founded in its knowledge base in the area.

Innovation policy as developed by the Dutch government (Innovatieplaform 2004, 2009) largely emphasizes the technological areas that our analysis shows the Netherlands to be strong in already. Chemistry, food & flowers are clearly present in Figure 1. Almost off necessity, the creative industries are not, since patents to works of art are not generally granted. This obviously reflects on the approach advocated here, but might equally give a clue as to what such a sector may be expected to contribute. Other domains such as electronics and computer technology are not a focus of innovation policy, while they seem to fit with the idea on which this policy is based of ‘backing winners’ (Nooteboom & Stam 2008). Our analysis thus suggests possible oversights of important technical domains, as well as possible (future) connections by the comprehensive empirical approach (Dolfsma 2009).

**The National Innovation System of India**. 936 Patents were granted in 2006 to parties located in India. Without throwing up a threshold of number of co-classifications any tie between two classes should have in order to enter the picture, the picture becomes quite difficult to understand, let alone interpret (as an illustration included as Fig. A1 in the appendix).

Introducing such a threshold of, for example, a *m*-slice of two (2), produces Figure 2. Here one finds that the Indian Innovation System's main strengths seem to be in chemistry, a relatively mature industry and knowledge field. Chemistry is a well-connected field in the Indian innovation system with that finds applications in food, cleaning and medical fields. Medical applications seem important, which may be where India's known strengths in production of generic medication shows (Chittoor *et al.* 2009). The set of companies involved in this industry, including Ranbaxy and Dr. Reddy's Laboratories, is reported to move into the phase of developing medication themselves, rather than copying medical innovations developed elsewhere. The industry is known to show a high propensity to patent, and so India can be expected to see its presence in patent databases enhanced in the years to come.

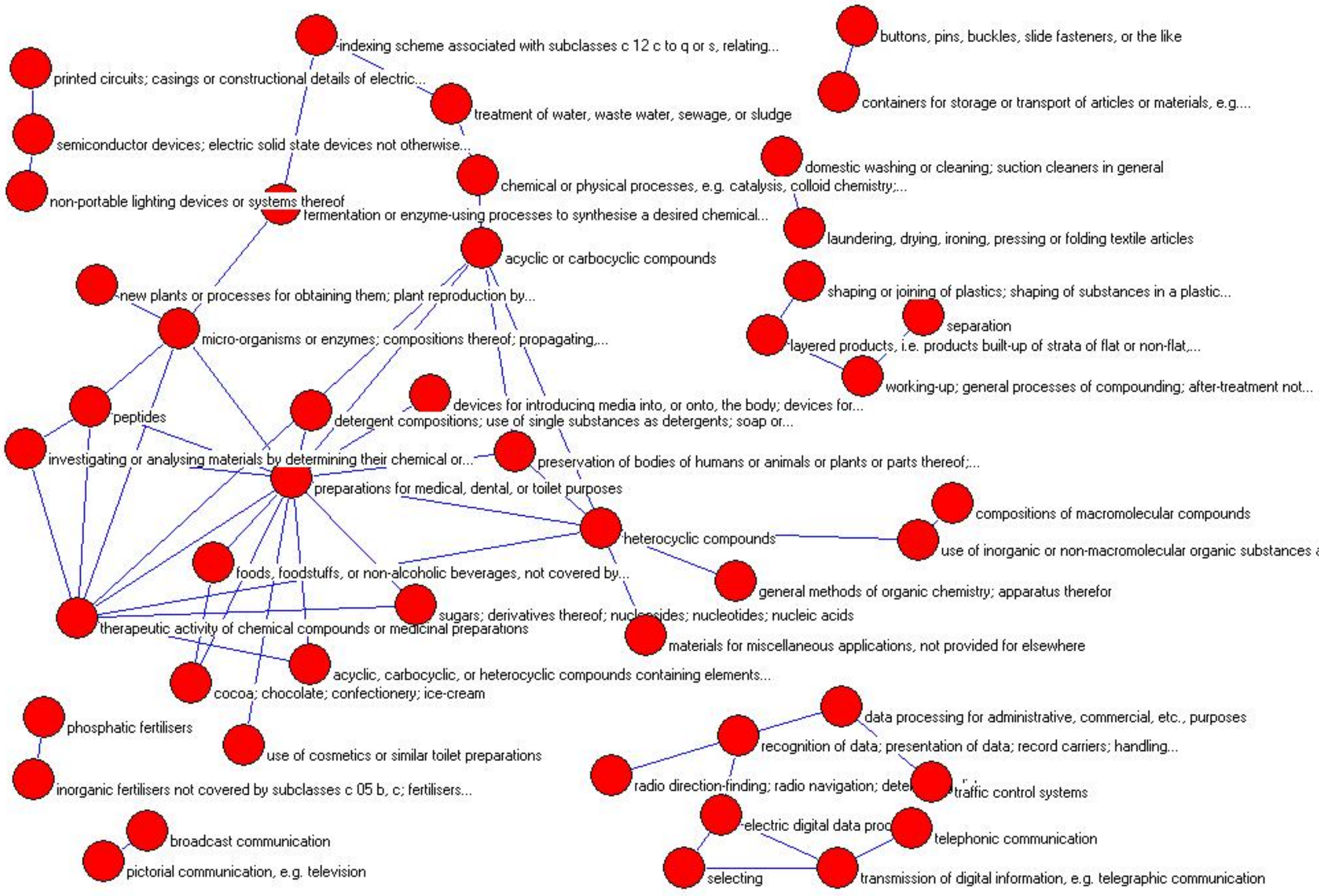

Figure 2: Patent classification categories and co-classification relations; core network for internationally registered Indian patents (*m*-slice = 2; 2006; *N* = 936). Source: WIPO.

What is striking is that another strong segment in the Indian economy, IT, is quite visibly present in the upper left and bottom right corners. Despite the fact that IT in India is mostly focused on services aspects, for which, of course, possibilities to apply for patents in most countries except for the US is restricted.

**Nanotechnology.** Patents have been granted to parties in India in the field of Nanotechnology (patent tag Y01N),[7] but these are filtered out very quickly for India when applying a *k*-core threshold. This is no surprise as nanotechnology is only recognized as a separate class recently. For India, it can be observed that nanotechnology only has a single tie with another patent class (‘Soil working in agriculture forestry etc.’).

Nanotechnology indeed is a separate field of knowledge development, at least from the perspective of application also when taking a sector innovation systems approach: an increasing number of patents are being granted. The approach to understanding innovation systems developed in this paper suggests how the field relates to other technological domains. Conceptually, thus, the approach advocated relates to the discussion of technological regimes (Dosi 1982) as it allows for a better understanding a domain through an analysis of its connections with adjacent domains. Using ideas developed in this paper for a sector or technological innovation system approach, it is possible to indicate which technological areas are close to nanotechnology field globally. When patents in any particular technological field are co-classified with another field one may expect knowledge to be exchanged between these classes. By repeating this analysis for the course of a number of years, the changing position of nanotechnology within the larger knowledge field can be indicated.

[7] Nanotechnology was recently added as an additional ‘tag’ to the existing database for all nano-technologies (Scheu *et al*. 2006).

Figure 3 pictures the Nanotech Innovation System without imposing an additional (*k*-core) threshold. One can thus conclude that the area of Nanotechnology is rather loosely connected to other technological regimes or fields, even though a process of fusion might be under way to become visible in patent applications some time from now (Islam & Miyazaki 2009). Based on our analysis, the way in which the technology is related to other fields of research does not show any structurally specific shape yet. While this may change in the future, it thus does appear to constitute a separate regime. Knowledge dynamics shaping this field has not yet developed to the extent that many industrial applications are to be expected, making such knowledge patentable. Knowledge development in the area of nanotechnology at present largely plays out in the scientific journals, where its dynamics might be of a very different nature (Leydesdorff 2008b).

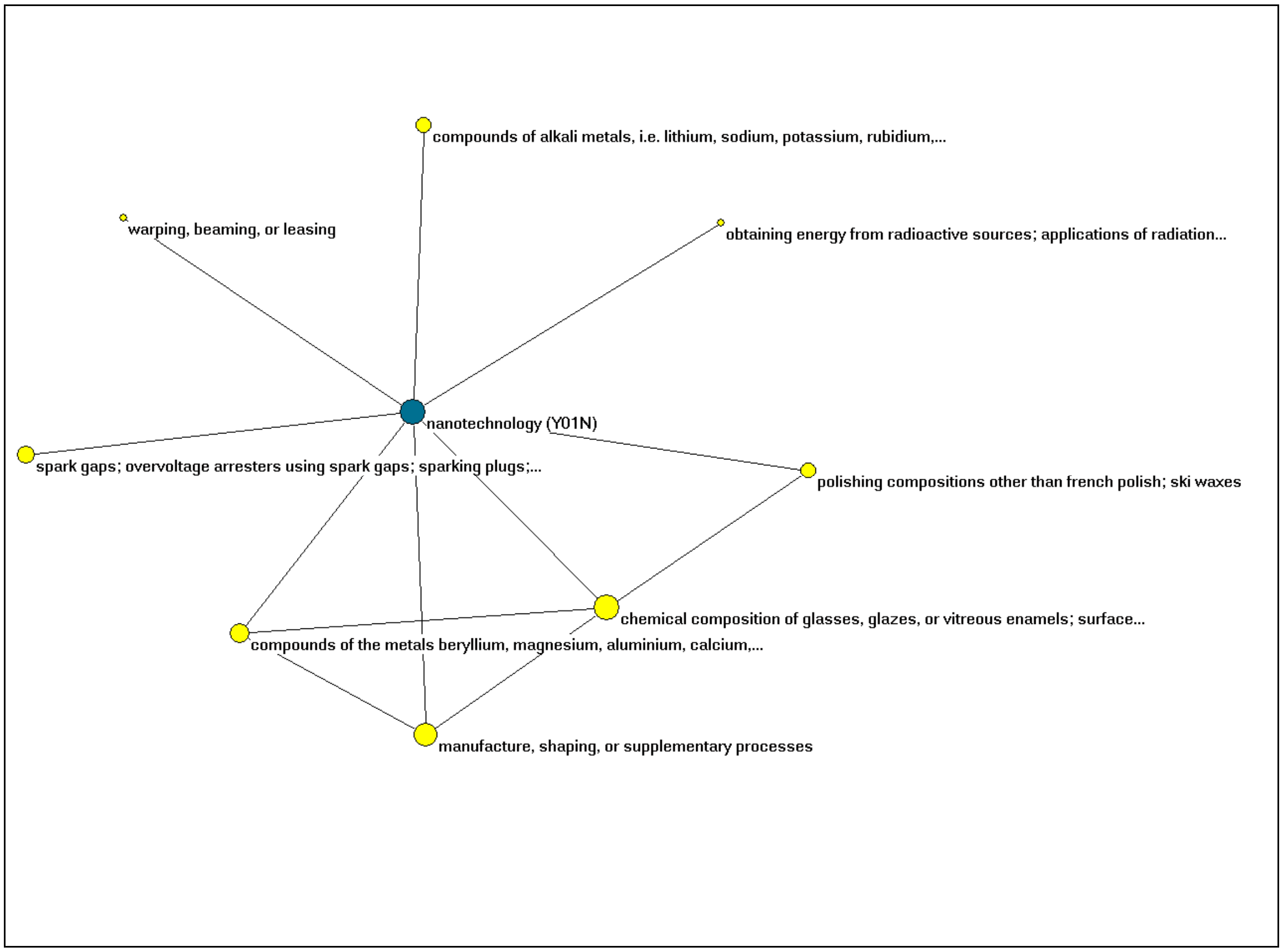

**Figure 3**: Technological Neighborhood for Nanotechnology, Y01N ($N$ = 762; $k$-core = 1; visualization based on the algorithm of Kamada & Kawai (1989).

## 4. Discussion and Conclusions

The literature on innovation systems has a tendency to get bogged down in case studies of the effects of specific institutions, for example on innovation in a specific countries, regions or specific sectors and technologies (Balzat & Hanusch 2004). Analyzing co-classification relations of patents granted by WIPO opens up avenues for research in the analysis of Systems of Innovation that promise to provide a more comprehensive perspective that is complementary with studies that have adopted a qualitative approach.

From a methodological point of view, acknowledging the drawbacks of patent data, discussed at length above, the approach suggested allows for a great deal of flexibility and rigor. From national point of view, which technological domains within a country are relatively stronger, as least in terms of patents applied for, becomes apparent at a glance. In addition, the way in which domains may interrelate in ways that may not be expected can be indicated as well. One may easily shift from a focus on national innovation systems to regional innovation systems. A social network approach using patent data can also provide insights into sectoral or technical innovation systems. It shows empirically where technical domains draw their knowledge from and where its knowledge is used. Developments in a specific technical domain may be plotted geographically as well, and can be animated over time (Leydesdorff & Rafols 2011). The approach suggested could thus be 'history friendly' (Nelson 1993) as well.

In this paper we have mainly done so in relation to national innovation systems, but similar analyses can be undertaken for sector and regional innovation systems. Making use of patent data in such a way has not been suggested before in innovation studies and so we offer three distinct contributions in this paper.

First, we offer a possible avenue for future research in the area of innovation studies and the field of NIS in particular that allows for comparison of innovation systems as well as a more comprehensive analysis of the development of a system over time.[8] Secondly, we show that widely available data on patents can be put to a different and more comprehensive use than has hitherto been done. This is a methodological advance of particular importance for innovation and industry studies. Thirdly, drawing on this, we offer empirical insights into important aspects of particular innovation systems, the Dutch and Indian innovation systems as well as the sector innovation system for nanotechnology, that might suggest some important policy implications as well. These insights can be replicated for any other nation or sector for which comparable data is available. Especially when studying developments over time, and taking into account the phase in a life cycle or stage of development that a region, country or sector is in, comparisons between these may have somewhat of a firmer base. These may be input for a better understanding of how specific institutions affect different aspects of innovation systems. Address details in patent data can allow for a better understanding of how the relations between actors help shape innovation systems (Cantner & Graf 2006).

[8] Compare the analysis of scientific journals over time (Leydesdorff & Schank 2008).

The picture that emerges for the Dutch innovation system, for example, is both familiar and somewhat surprising. What is to be expected is the strong presence of the electronics, computer, and optical clusters. Internationally well-recognized and established industrial firms such as electronics giant Philips, world leader in semi-conductor productions ASML, or producer of copy machines Océ feature in this corner of the innovation system. What is more surprising is the strongly intertwined chemical, biotechnical and pharmaceutical cluster, especially with application for (veterinary) medication and (functional) food. The presence of this element in the Dutch NIS is not generally recognized, and, given the high propensity to patent in the related industries (Arundel & Kabla 1998), would suggest that the Dutch innovation becomes increasingly dependent on intellectual property law. For India, the picture may be more surprising even. IT, which in India is known to be a strong sector, is noticeably absent, while chemistry and pharmacy are most strongly present. Looking at nanotech, globally, taking a sector innovation systems approach, it appear that it is very much a technology in development, currently rather loosely connected to the broader set of technological fields as currently recognized.

## Appendix

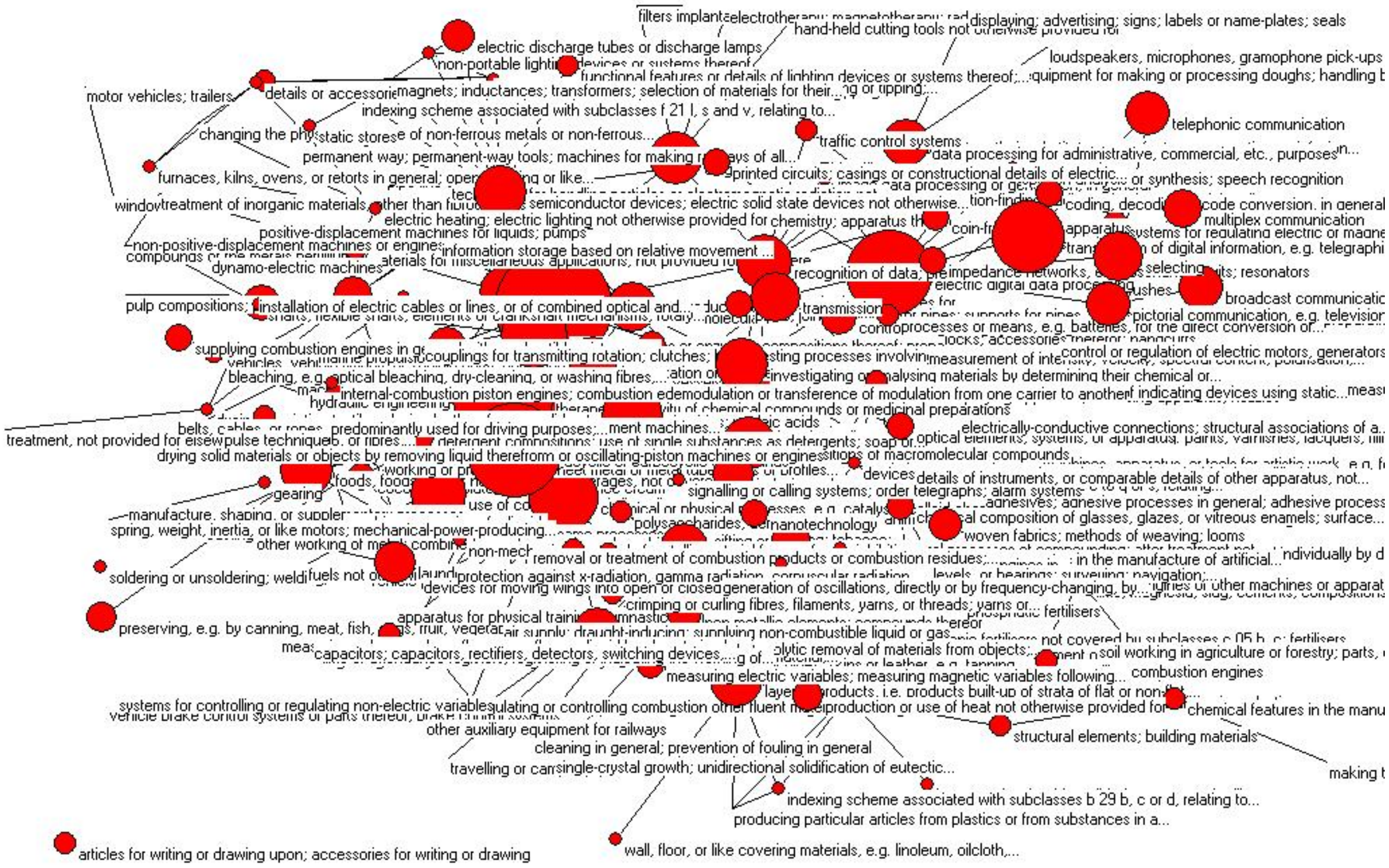

Figure A1: Patent classification categories and co-classification relations for India, 2006.

(*N* = 936; *k*-core = 1) Source: WIPO.